\documentclass[aps,prl,preprint,superscriptaddress,showpacs]{revtex4-1}
\usepackage{graphics}
\begin{document}

\title{Femtosecond dynamics of the collinear-to-spiral antiferromagnetic phase transition in CuO}


\author{S. L. Johnson}
\affiliation{Swiss Light Source, Paul Scherrer Institut, CH-5232 Villigen PSI, Switzerland}
\altaffiliation{Current address:  Physics Department, ETH Zurich, 8093 Zurich, Switzerland}
\author{R. A. de Souza}
\author{U. Staub}
\author{P. Beaud}
\author{E. M\"{o}hr-Vorobeva}
\author{G. Ingold}
\author{A. Caviezel}
\author{V. Scagnoli}
\affiliation{Swiss Light Source, Paul Scherrer Institut, CH-5232 Villigen PSI, Switzerland}

\author{W. F. Schlotter}
\author{J. J. Turner}
\affiliation{The Linac Coherent Light Source, SLAC National Accelerator Laboratory, Menlo Park, CA 94025 USA}

\author{O. Krupin}
\affiliation{European XFEL GmbH Albert Einstein Ring 19, 22 607 Hamburg, Germany}
\affiliation{The Linac Coherent Light Source, SLAC National Accelerator Laboratory, Menlo Park, CA 94025 USA}

\author{W.-S. Lee}
\affiliation{SIMES, SLAC National Accelerator Laboratory and Stanford University, Menlo Park, CA 94305, USA}
\affiliation{SSRL, SLAC National Accelerator Laboratory, Menlo Park, CA 94305, USA}

\author{Y.-D. Chuang}
\affiliation{Lawrence Berkeley National Laboratory, Berkeley, CA 94720, USA}

\author{L. Patthey}
\affiliation{Swiss Light Source, Paul Scherrer Institut, CH-5232 Villigen PSI, Switzerland}

\author{R. G. Moore}
\affiliation{SIMES, SLAC National Accelerator Laboratory and Stanford University, Menlo Park, CA 94305, USA}

\author{D. Lu}
\affiliation{SSRL, SLAC National Accelerator Laboratory, Menlo Park, CA 94305, USA}

\author{M. Yi}
\affiliation{SIMES, SLAC National Accelerator Laboratory and Stanford University, Menlo Park, CA 94305, USA}

\author{P. S. Kirchmann}
\affiliation{SIMES, SLAC National Accelerator Laboratory and Stanford University, Menlo Park, CA 94305, USA}

\author{M. Trigo}
\affiliation{PULSE, SLAC National Accelerator Laboratory, Menlo Park, CA 94025 USA}

\author{P. Denes}
\author{Dionisio Doering}
\affiliation{Lawrence Berkeley National Laboratory, Berkeley, CA 94720, USA}

\author{Z. Hussain}
\affiliation{Lawrence Berkeley National Laboratory, Berkeley, CA 94720, USA}

\author{Z.-X. Shen}
\affiliation{SIMES, SLAC National Accelerator Laboratory and Stanford University, Menlo Park, CA 94305, USA}

\author{D. Prabhakaran}
\author{A. T. Boothroyd}
\affiliation{Department of Physics, University of Oxford, Clarendon Laboratory, Parks Road, Oxford OX1 3PU, UK}

\begin{abstract}
We report on the ultrafast dynamics of magnetic order in a single crystal of CuO at a temperature of 207~K in response to strong optical excitation using femtosecond resonant x-ray diffraction.  In the experiment, a femtosecond laser pulse induces a sudden, nonequilibrium increase in magnetic disorder.  After a short delay ranging from 400~fs to 2~ps, we observe changes in the relative intensity of the magnetic ordering diffraction peaks that indicate a shift from a collinear commensurate phase to a spiral incommensurate phase.  These results indicate that the ultimate speed for this antiferromagnetic re-orientation transition in CuO is limited by the long-wavelength magnetic excitation connecting the two phases. 
\end{abstract}


\maketitle

The ability to manipulate magnetic order has become vital to modern data storage and processing technology.  As this technology is pushed to handle ever increasing data rates, fundamental questions arise on the ultimate time scales required for changes in magnetic order and how it is possible to induce them in a controlled way~\cite{Kirilyuk:2010ha,Tudosa:2004uq,Kimel:2004by,Stanciu:2007tu,Bigot:2009uh,Holldack:2010un}.  In several ferromagnetic materials, experiments have demonstrated demagnetization on timescales of several hundred femtoseconds~\cite{Beaurepaire:1996vo,Scholl:1997to,Rhie:2003tq,Koopmans:2009ul,Stamm:2007hy}.  While this addresses one aspect of the problem, another potential focus is the dynamics of phase transitions between different types of magnetic order~\cite{Ju:2004jw,Radu:2010ex,Radu:2011kr,Seu:2010ec}.  Of particular interest here are phase transitions involving antiferromagnetic spiral spin structures resulting from magnetic frustration, which often lead to multiferroicity~\cite{Cheong:2007gd}.  
CuO, well-known as a starting material for high-T$_c$ cuprate superconductors, shows one important example of such a phase transition with some unusual characteristics.  


The atomic structure of CuO can be described as connected networks of zigzag Cu-O chains oriented along the [$1$\,$0$\,$-1$] and [$1$\,$0$\,$1$] directions~\cite{Asbrink:1970vb}.  Depending on the temperature, the spins on the Cu atoms take on different ordering patterns~\cite{Forsyth:1988tv,Yang:1989uy,Brown:1991uz}.
Figure~1 shows the structure of the magnetically ordered phases of CuO, projected into the ($0$\,$1$\,$0$) plane.  The low temperature phase has a collinear antiferromagnetic structure.  It has a commensurate (CM) ordering wave-vector (0.5 0 -0.5) that forms a magnetic unit cell twice the dimensions of the structural unit cell in both the $\mathbf{a}$ and $\mathbf{c}$ directions.  A multiferroic phase appears in the temperature range $213-230$~K~\cite{Kimura:2008ks}.  In this phase the magnetic order is qualitatively different:  it is a non-collinear antiferromagnet with an incommensurate (ICM) ordering wave-vector ($0.506$\,\,$0$\,\,$-0.483$) that is not pinned to the underlying lattice periodicity.  The spiral spin structure of this phase results in a small ferroelectric moment.  Interestingly, the temperature at which this induced multiferroicity exists in CuO is much higher than has been observed in other materials, a consequence of the unusually large magnetic superexchange interactions in copper oxides that may also have a connection to the high-T$_c$ superconductivity often observed in these compounds~\cite{Kimura:2008ks}.
At temperatures above 230~K CuO loses long-range 3-D magnetic order.   

These changes in magnetic order are the result of frustration between ferromagnetic  and antiferromagnetic exchange interactions among the spins on the Cu atoms.  The strongest coupling in the system is a nearest-neighbor antiferromagnetic (AF) interaction along the $[$1$\,$0$\,$-1$]$ direction, with exchange interaction $J_{AF} = 74$~meV~\cite{Ain:1989uk,Boothroyd:1997wi}.  The large strength of this interaction results in 1-D collinear AF chains along [$1$\,$0$\,$-1$] in both the CM and ICM phases.  To explain the CM/ICM transition, Yablonskii~\cite{Yablonskii:1990te} proposed a model that considers the energy $E$ from effective nearest-neighbor and next-nearest-neighbor interactions among these AF chains:
\begin{equation}
E = \sum_n [J_1 \mathbf{S}_n \mathbf{S}_{n+1}+J_2 \mathbf{S}_n \mathbf{S}_{n+2} + I (\mathbf{S}_n \mathbf{S}_{n+1}) (\mathbf{S}_{n+1} \mathbf{S}_{n+2})]\label{eq:E}
\end{equation}
where $\mathbf{S}_n$ refers to the spin of a Cu atom along the c axis with index $n$ as shown in figure~1,  
$J_1 < 0$ is a nearest-neighbor ferromagnetic exchange interaction, $J_2 > 0$ is a next-nearest-neighbor antiferromagnetic exchange interaction, and the biquadratic exchange interaction $I>0$ is assumed to be small relative to the others.  
The biquadratic interaction term acts to stabilize the CM phase at low temperature.  By invoking a mean-field approximation and taking $S$ as the spatially averaged value of the spin magnitude, the CM phase has a lower energy for $S^2 > J_1^2 / 8 I J_2$.  Since $S$ decreases with temperature, at some point the value of $S^2$ drops below this threshold and the ICM phase becomes more stable.

Although this particular role of a biquadratic energy term as the mechanism for the low-T stability of the CM phase may be unique to CuO, the role of thermal disorder in the phase transition is more general.  In a mean-field treatment, temperature is essentially modeled as a change in the average value of an order parameter due to disorder.   In multiferroic and other materials with competing interactions and order parameters this can result in order-order phase transitions.  CuO presents one particular example of how this disorder can lead to a qualitative change in the nature of magnetic order.

Temperature is only one way to control the level of spin disorder and thus the value of $S$.  Alternatively, optical excitation of CuO above the charge transfer bandgap at 1.4~eV~\cite{Sukhorukov:1995ts} can directly induce non-thermal spin disorder by exciting electron transfer.  This allows us to use an intense, short pulse of light to nearly instantaneously drive the magnetic order toward the ICM phase.  We can then explore the limiting time scales of the transition itself.


We use femtosecond time-resolved resonant x-ray scattering to observe the effect of optical excitation on the CM and ICM magnetic structures.  For x-rays with an energy tuned near the Cu L$_3$ edge at 930~eV, the scattering from magnetic contributions is strongly enhanced.  The diffracted intensity at the magnetic ordering wavevector has no measurable structural component, but is instead a direct measure of the ordered component of the dipole magnetic moment and an orbital flux (toroidal moment) which coexist within the CuO plaquette~\cite{Scagnoli:2011db,Wu:2010cf}. The orbital flux component is an important ingredient in the description of the pseudo-gap phase order parameter in cuprate superconductors~\cite{Varma:1997ue}. By using $< 10$ fs pulses of x-rays in combination with synchronized 40~fs pulses of 800~nm light, we investigate the magnetic dynamics using a pump-probe method.

The sample is a single crystal of CuO grown from high-purity CuO powder using the floating-zone technique~\cite{Prabhakaran:2003hd}.  Cylindrical rods of length 10 cm and diameter 12 mm were sintered for a three day period in a vertical furnace at 950$^\circ$ under oxygen. The crystal was then grown in an optical floating-zone furnace using a mixed oxygen/argon atmosphere.  
The surface was miscut at an angle of 10$^\circ$ with respect to the (0.5 0 -0.5) planes.

The pump-probe resonant x-ray diffraction measurements were performed at the Linac Coherent Light Source (LCLS)~\cite{Emma:2010gu}, using the soft x-ray materials science (SXR) instrument.    The crystal was cooled to a temperature of 207~K in vacuum using a He-flow cryostat.  
To excite the sample, a 40~fs full-width-at-half maximum pulse of 800~nm wavelength light hit the sample with an horizontal angle of 64.1$^\circ$ and $\pi$-polarization with respect to the (0.5 0 -0.5) planes.  
Sub-10 fs pulses of x-rays probed the magnetic ordering.  For these measurements the photon energy was tuned to optimize output at 930~eV.  A variable-line-spaced plane-grating monochromator set this as the x-ray energy to a bandwidth of 1.2~eV.  A set of Kirkpatrick-Baez mirrors focused the x-ray beam to a 300~micron diameter spot on the sample.  The incident x-ray beam was collinear with the excitation beam and $\pi$-polarized with respect to the (0.5 0 -0.5) planes.  
A continuously moving optical delay line in the path of the pump beam set the relative delay of the two synchronized pulses.  For each pulse, this timing was corrected 
by electron bunch time arrival monitors.

A custom-built CCD camera detected the diffracted x-rays.  The camera is able to isolate individual x-ray pulses from the 60~Hz repetition rate of the LCLS.  The sample angle was tuned intermediate between the optimal Bragg conditions for the CM and ICM phases, which are separated by less than $1.5^\circ$.  This allowed simultaneous measurement of both peaks.  The peak positions did not change as a function of time delay, indicating that changes in the normal component of the reciprocal lattice vectors (due to, for example, lattice expansion) 
do not contribute to the measured data.  
To correct for intensity and energy fluctuations of the source, the overall image intensities were normalized by the drain current signal from a thin Al-backed foil placed in the x-ray beam before the sample.  The intensity of the peaks for a given image was estimated by summing the image data along the short axis of the CCD image and then fitting the resulting curves to the sum of two Gaussians.  The integral of each Gaussian was then taken as the intensity.  

Figure~2 shows images of the magnetic diffraction signal at times before and after excitation.  The two peaks show contributions from each of the two types of magnetic ordering.  They coexist in different domains at this temperature due to the first-order nature of the phase transition.  Because the estimated probe depth of 50~nm 
is less than the correlation length (approximately 100--200~nm), 
the measured domains are distributed predominantly along the surface of the sample.  The majority of these domains are initially in the CM phase.  After excitation, the intensities of both peaks decrease but the relative intensity shifts in favor of the ICM phase, indicating that the optically-induced spin disorder has indeed changed the average magnetic order.

Figure~3a shows the time-dependence of the intensities for a fluence of 28~mJ/cm$^2$.  When the optical pulse hits the sample, both peaks show a sudden decrease over a time scale of 300~fs.  The relative change in intensity here is the same for both peaks; we also observe this for other excitation fluences.  This sudden drop has a simple interpretation:  it is a direct measure of the additional spin disorder in the system induced by optical excitation.  This is analogous to the drop in intensity observed in structural diffraction peaks due to non-thermal lattice disorder~\cite{Johnson:2009ty,Johnson:2010em}.    
The fast speed indicates that the disordering occurs on a time scale faster than the estimated experimental time resolution of 300~fs.  The exact mechanism of the laser-induced spin disorder is not unambiguously determined from the experimental results.

At a time slightly later than the initial drop, the time dependence of the two peaks begins to differ.  Figure~3b shows the time-dependence of the ratio between the ICM and CM intensities.  These data show the relative contributions of each phase as a function of time after the onset of spin disorder.  The curve superimposed on the data is a fit to a bi-exponential curve that begins at a delayed time $t_{p}$.  As shown in the inset to figure~3b, the fitted value of $t_{p}$ is significantly larger than zero for all measured excitation fluences.  
The delay is 2~ps for low excitation levels and becomes shorter with increasing fluence until reaching a ``saturation'' level of approximately 400~fs.  Note that since $t_p$ is derived from a comparison of mathematically independent data sets, it is not limited by the time resolution of the experiment. 
After $t_p$, the relative intensity of the incommensurate peak grows. 
Initially there is a fast component with a time constant of $1.1\pm0.3\,\textrm{ps}$  for the lowest fluence of 7~mJ/cm$^2$, increasing slightly with fluence to a value of $1.86\pm 0.14$~ps at $39$~mJ/cm$^2$.  There is also a much slower time scale component with a time constant ranging from approximately 35--60 ps, becoming faster with increasing fluence.  In this context it is important to note that these fits considered only times up to 20 ps after excitation;  the later time behavior is not quantitatively well described by this model.  
The ratio at 100~ps increases with excitation fluence~\cite{EPAPS}.  At the maximum fluence of 39~mJ/cm$^2$ we observed an increase of the ICM/CM ratio by a factor of 2.8 over the initial ratio.  

In principle, there are at least two possible explanations for the difference in the time response between the two phases.  One possibility is that the relaxation processes for the two magnetic phases differ intrinsically.  The dynamics can arise from both changes in the orbital flux and the magnetic moment contributions to the magnetic scattering.  Although orbital current is expected to couple more strongly to the lattice since it is related to inversion symmetry breaking, we expect that the much stronger interactions between magnetic dipole moments control the dynamics in CuO.
The magnetic dipole disorder induced by optical excitation of the charge-transfer band is expected to create predominantly local exchange of spins among Cu atoms, resulting in a broad spectrum of magnetic excitations throughout the Brillouin zone.  Interactions of these magnetic excitations both with each other and with the lattice and charge degrees of freedom will at some point result in a thermalized distribution of magnetic excitations in equilibrium with other energy subsystems in the crystal.  Since the initial steps of this process involve relaxation of high-frequency magnetic excitations ($h \nu \sim 150$~meV)~\cite{Boothroyd:1997wi}, it is not offhand clear why there should be a difference between two phases that mostly differ in their low-frequency, long-wavelength magnetic behavior.  
Measurements (see~\cite{EPAPS}) of the time-dependent intensity of the CM peak at 150~K where no ICM peak was observed show behavior that is qualitatively different:  the intensity of the CM peak \textit{increases} over a time scale of 2-3~ps.  It thus appears that the decrease in the CM peak intensity observed closer to the phase transition temperature is strongly related to the transition itself.  

Alternatively, the difference in the time response can be the result of a shift in population of the CM to the ICM phase, possibly starting at the boundaries between pre-existing domains.
The excitation-dependent delay may be explained qualitatively in terms of the configuration schematics in figure~4.  In general, any change in structure can be described as motion along a set of coordinates associated with normal excitation modes.  Although in fact the set of magnetic coordinates that connect the CM and ICM phases involves several different modes, for the sake of simplicity we consider only changes along one coordinate that physically corresponds to an acoustic-branch magnetic excitation with wavevector $q = q_\textrm{ICM}-q_\textrm{CM} \approx 0$.  This is a useful approximation since this is likely to be among the lowest frequency modes that mediate the change in magnetic order.  Initially, the energy of the CM phase is lower than that of the ICM phase.  The energy barrier between the CM and ICM phases may be roughly estimated as the magnitude of the biquadratic term in equation~\ref{eq:E}, evaluated as $IS^4$ in the mean field approximation for the CM phase.  At low excitation fluences, the relative energies of the two phases shifts in favor of the ICM phase.  The energy barrier is, however, still sizable.  The system requires a non-zero amount of time to hop over the energy barrier and enter into the new phase.  As the excitation fluence increases, the size of the energy barrier decreases until it reaches a level where it no longer appreciably impedes the speed of the transition.  At these very high excitation levels, we might expect that the time scale is limited purely by dynamics:  the time required for a coherent, $1/4$ oscillation of a low momentum $q\approx 0$ spin wave.  
Inelastic neutron scattering measurements in the CM phase have estimated a period of 1.6~ps for such excitations~\cite{Yang:1989uy}.  Taking 1/4 of this period yields 400~fs, in agreement with the delay observed at the highest excitation fluences. This is analogous to the situation with structural phase transitions, where time scales of a fraction of a phonon period are frequently cited as minimum time scales~\cite{Cavalleri:2004eh}.  

After $t_p$, the fast increase of the ICM/CM ratio can be considered as the first step of the actual phase transition.  
The subsequent increase of the ratio over several tens of picoseconds occurs at approximately time as a slow decrease in the overall diffracted intensity.  We tentatively interpret this as an intermediate step in thermalization that further suppresses the spatially averaged $S^2$ and pushes the system even more toward the ICM phase.   

The concept of using intense optical pulses to control magnetic order has the potential for a wide variety of applications.
We have shown here one possible way to accomplish a partial, transient change in average magnetic order from one type to another by using a 40~fs duration optical pulse to initiate the process.  In addition, we have identified the limiting time scales for how fast the first steps of such a change may occur.  The observed changes in magnetic order may be determined by the growth of pre-existing domains;  further studies with improved spatial resolution 
may help to clarify this.  







\begin{acknowledgments}
This research was carried out on the SXR Instrument at
the LCLS, a division of SLAC and an Office of Science user facility operated
by Stanford University for the U.S. Department of Energy. The SXR
Instrument is funded by a consortium including the
LCLS, Stanford University through SIMES, LBNL,
the University of Hamburg through the BMBF priority program FSP 301, and
the Center for Free Electron Laser Science (CFEL).  This work was supported by the NCCR-MUST and NCCR-MaNEP,
research instruments of the Swiss National Science Foundation.  P.S.K.
acknowledges support by the 
Alexander-von-Humboldt Foundation through a Feodor-Lynen scholarship.
\end{acknowledgments}

%


\begin{thebibliography}{35}%
\makeatletter
\providecommand \@ifxundefined [1]{%
 \@ifx{#1\undefined}
}%
\providecommand \@ifnum [1]{%
 \ifnum #1\expandafter \@firstoftwo
 \else \expandafter \@secondoftwo
 \fi
}%
\providecommand \@ifx [1]{%
 \ifx #1\expandafter \@firstoftwo
 \else \expandafter \@secondoftwo
 \fi
}%
\providecommand \natexlab [1]{#1}%
\providecommand \enquote  [1]{``#1''}%
\providecommand \bibnamefont  [1]{#1}%
\providecommand \bibfnamefont [1]{#1}%
\providecommand \citenamefont [1]{#1}%
\providecommand \href@noop [0]{\@secondoftwo}%
\providecommand \href [0]{\begingroup \@sanitize@url \@href}%
\providecommand \@href[1]{\@@startlink{#1}\@@href}%
\providecommand \@@href[1]{\endgroup#1\@@endlink}%
\providecommand \@sanitize@url [0]{\catcode `\\12\catcode `\$12\catcode
  `\&12\catcode `\#12\catcode `\^12\catcode `\_12\catcode `\%12\relax}%
\providecommand \@@startlink[1]{}%
\providecommand \@@endlink[0]{}%
\providecommand \url  [0]{\begingroup\@sanitize@url \@url }%
\providecommand \@url [1]{\endgroup\@href {#1}{\urlprefix }}%
\providecommand \urlprefix  [0]{URL }%
\providecommand \Eprint [0]{\href }%
\providecommand \doibase [0]{http://dx.doi.org/}%
\providecommand \selectlanguage [0]{\@gobble}%
\providecommand \bibinfo  [0]{\@secondoftwo}%
\providecommand \bibfield  [0]{\@secondoftwo}%
\providecommand \translation [1]{[#1]}%
\providecommand \BibitemOpen [0]{}%
\providecommand \bibitemStop [0]{}%
\providecommand \bibitemNoStop [0]{.\EOS\space}%
\providecommand \EOS [0]{\spacefactor3000\relax}%
\providecommand \BibitemShut  [1]{\csname bibitem#1\endcsname}%
\let\auto@bib@innerbib\@empty
\bibitem [{\citenamefont {Kirilyuk}\ \emph {et~al.}(2010)\citenamefont
  {Kirilyuk}, \citenamefont {Kimel},\ and\ \citenamefont
  {Rasing}}]{Kirilyuk:2010ha}%
  \BibitemOpen
  \bibfield  {author} {\bibinfo {author} {\bibfnamefont {A.}~\bibnamefont
  {Kirilyuk}}, \bibinfo {author} {\bibfnamefont {A.}~\bibnamefont {Kimel}}, \
  and\ \bibinfo {author} {\bibfnamefont {T.}~\bibnamefont {Rasing}},\
  }\href@noop {} {\bibfield  {journal} {\bibinfo  {journal} {Rev. Mod. Phys.}\
  }\textbf {\bibinfo {volume} {82}},\ \bibinfo {pages} {2731} (\bibinfo {year}
  {2010})}\BibitemShut {NoStop}%
\bibitem [{\citenamefont {Tudosa}\ \emph {et~al.}(2004)\citenamefont {Tudosa},
  \citenamefont {Stamm}, \citenamefont {Kashuba}, \citenamefont {King},
  \citenamefont {Siegmann}, \citenamefont {St{\"o}hr}, \citenamefont {Ju},
  \citenamefont {Lu},\ and\ \citenamefont {Weller}}]{Tudosa:2004uq}%
  \BibitemOpen
  \bibfield  {author} {\bibinfo {author} {\bibfnamefont {I.}~\bibnamefont
  {Tudosa}}, \bibinfo {author} {\bibfnamefont {C.}~\bibnamefont {Stamm}},
  \bibinfo {author} {\bibfnamefont {A.~B.}\ \bibnamefont {Kashuba}}, \bibinfo
  {author} {\bibfnamefont {F.}~\bibnamefont {King}}, \bibinfo {author}
  {\bibfnamefont {H.~C.}\ \bibnamefont {Siegmann}}, \bibinfo {author}
  {\bibfnamefont {J.}~\bibnamefont {St{\"o}hr}}, \bibinfo {author}
  {\bibfnamefont {G.}~\bibnamefont {Ju}}, \bibinfo {author} {\bibfnamefont
  {B.}~\bibnamefont {Lu}}, \ and\ \bibinfo {author} {\bibfnamefont
  {D.}~\bibnamefont {Weller}},\ }\href@noop {} {\bibfield  {journal} {\bibinfo
  {journal} {Nature}\ }\textbf {\bibinfo {volume} {428}},\ \bibinfo {pages}
  {831} (\bibinfo {year} {2004})}\BibitemShut {NoStop}%
\bibitem [{\citenamefont {Kimel}\ \emph {et~al.}(2004)\citenamefont {Kimel},
  \citenamefont {Kirilyuk}, \citenamefont {Tsvetkov}, \citenamefont {Pisarev},\
  and\ \citenamefont {Rasing}}]{Kimel:2004by}%
  \BibitemOpen
  \bibfield  {author} {\bibinfo {author} {\bibfnamefont {A.~V.}\ \bibnamefont
  {Kimel}}, \bibinfo {author} {\bibfnamefont {A.}~\bibnamefont {Kirilyuk}},
  \bibinfo {author} {\bibfnamefont {A.}~\bibnamefont {Tsvetkov}}, \bibinfo
  {author} {\bibfnamefont {R.~V.}\ \bibnamefont {Pisarev}}, \ and\ \bibinfo
  {author} {\bibfnamefont {T.}~\bibnamefont {Rasing}},\ }\href@noop {}
  {\bibfield  {journal} {\bibinfo  {journal} {Nature}\ }\textbf {\bibinfo
  {volume} {429}},\ \bibinfo {pages} {850} (\bibinfo {year}
  {2004})}\BibitemShut {NoStop}%
\bibitem [{\citenamefont {Stanciu}\ \emph {et~al.}(2007)\citenamefont
  {Stanciu}, \citenamefont {Hansteen}, \citenamefont {Kimel}, \citenamefont
  {Kirilyuk}, \citenamefont {Tsukamoto}, \citenamefont {Itoh},\ and\
  \citenamefont {Rasing}}]{Stanciu:2007tu}%
  \BibitemOpen
  \bibfield  {author} {\bibinfo {author} {\bibfnamefont {C.~D.}\ \bibnamefont
  {Stanciu}}, \bibinfo {author} {\bibfnamefont {F.}~\bibnamefont {Hansteen}},
  \bibinfo {author} {\bibfnamefont {A.~V.}\ \bibnamefont {Kimel}}, \bibinfo
  {author} {\bibfnamefont {A.}~\bibnamefont {Kirilyuk}}, \bibinfo {author}
  {\bibfnamefont {A.}~\bibnamefont {Tsukamoto}}, \bibinfo {author}
  {\bibfnamefont {A.}~\bibnamefont {Itoh}}, \ and\ \bibinfo {author}
  {\bibfnamefont {T.}~\bibnamefont {Rasing}},\ }\href@noop {} {\bibfield
  {journal} {\bibinfo  {journal} {Phys. Rev. Lett.}\ }\textbf {\bibinfo
  {volume} {99}},\ \bibinfo {pages} {047601} (\bibinfo {year}
  {2007})}\BibitemShut {NoStop}%
\bibitem [{\citenamefont {Bigot}\ \emph {et~al.}(2009)\citenamefont {Bigot},
  \citenamefont {Vomir},\ and\ \citenamefont {Beaurepaire}}]{Bigot:2009uh}%
  \BibitemOpen
  \bibfield  {author} {\bibinfo {author} {\bibfnamefont {J.-Y.}\ \bibnamefont
  {Bigot}}, \bibinfo {author} {\bibfnamefont {M.}~\bibnamefont {Vomir}}, \ and\
  \bibinfo {author} {\bibfnamefont {E.}~\bibnamefont {Beaurepaire}},\
  }\href@noop {} {\bibfield  {journal} {\bibinfo  {journal} {Nature Phys.}\
  }\textbf {\bibinfo {volume} {5}},\ \bibinfo {pages} {515} (\bibinfo {year}
  {2009})}\BibitemShut {NoStop}%
\bibitem [{\citenamefont {Holldack}\ \emph {et~al.}(2010)\citenamefont
  {Holldack}, \citenamefont {Pontius}, \citenamefont {Schierle}, \citenamefont
  {Kachel}, \citenamefont {Soltwisch}, \citenamefont {Mitzner}, \citenamefont
  {Quast}, \citenamefont {Springholz},\ and\ \citenamefont
  {Weschke}}]{Holldack:2010un}%
  \BibitemOpen
  \bibfield  {author} {\bibinfo {author} {\bibfnamefont {K.}~\bibnamefont
  {Holldack}}, \bibinfo {author} {\bibfnamefont {N.}~\bibnamefont {Pontius}},
  \bibinfo {author} {\bibfnamefont {E.}~\bibnamefont {Schierle}}, \bibinfo
  {author} {\bibfnamefont {T.}~\bibnamefont {Kachel}}, \bibinfo {author}
  {\bibfnamefont {V.}~\bibnamefont {Soltwisch}}, \bibinfo {author}
  {\bibfnamefont {R.}~\bibnamefont {Mitzner}}, \bibinfo {author} {\bibfnamefont
  {T.}~\bibnamefont {Quast}}, \bibinfo {author} {\bibfnamefont
  {G.}~\bibnamefont {Springholz}}, \ and\ \bibinfo {author} {\bibfnamefont
  {E.}~\bibnamefont {Weschke}},\ }\href@noop {} {\bibfield  {journal} {\bibinfo
   {journal} {Appl. Phys. Lett.}\ }\textbf {\bibinfo {volume} {97}},\ \bibinfo
  {pages} {062502} (\bibinfo {year} {2010})}\BibitemShut {NoStop}%
\bibitem [{\citenamefont {Beaurepaire}\ \emph {et~al.}(1996)\citenamefont
  {Beaurepaire}, \citenamefont {Merle}, \citenamefont {Daunois},\ and\
  \citenamefont {Bigot}}]{Beaurepaire:1996vo}%
  \BibitemOpen
  \bibfield  {author} {\bibinfo {author} {\bibfnamefont {E.}~\bibnamefont
  {Beaurepaire}}, \bibinfo {author} {\bibfnamefont {J.-C.}\ \bibnamefont
  {Merle}}, \bibinfo {author} {\bibfnamefont {A.}~\bibnamefont {Daunois}}, \
  and\ \bibinfo {author} {\bibfnamefont {J.-Y.}\ \bibnamefont {Bigot}},\
  }\href@noop {} {\bibfield  {journal} {\bibinfo  {journal} {Phys. Rev. Lett.}\
  }\textbf {\bibinfo {volume} {76}},\ \bibinfo {pages} {4250} (\bibinfo {year}
  {1996})}\BibitemShut {NoStop}%
\bibitem [{\citenamefont {Scholl}\ \emph {et~al.}(1997)\citenamefont {Scholl},
  \citenamefont {Baumgarten}, \citenamefont {Jacquemin},\ and\ \citenamefont
  {Eberhardt}}]{Scholl:1997to}%
  \BibitemOpen
  \bibfield  {author} {\bibinfo {author} {\bibfnamefont {A.}~\bibnamefont
  {Scholl}}, \bibinfo {author} {\bibfnamefont {L.}~\bibnamefont {Baumgarten}},
  \bibinfo {author} {\bibfnamefont {R.}~\bibnamefont {Jacquemin}}, \ and\
  \bibinfo {author} {\bibfnamefont {W.}~\bibnamefont {Eberhardt}},\ }\href@noop
  {} {\bibfield  {journal} {\bibinfo  {journal} {Phys. Rev. Lett.}\ }\textbf
  {\bibinfo {volume} {79}},\ \bibinfo {pages} {5146} (\bibinfo {year}
  {1997})}\BibitemShut {NoStop}%
\bibitem [{\citenamefont {Rhie}\ \emph {et~al.}(2003)\citenamefont {Rhie},
  \citenamefont {D{\"u}rr},\ and\ \citenamefont {Eberhardt}}]{Rhie:2003tq}%
  \BibitemOpen
  \bibfield  {author} {\bibinfo {author} {\bibfnamefont {H.~S.}\ \bibnamefont
  {Rhie}}, \bibinfo {author} {\bibfnamefont {H.~A.}\ \bibnamefont {D{\"u}rr}},
  \ and\ \bibinfo {author} {\bibfnamefont {W.}~\bibnamefont {Eberhardt}},\
  }\href@noop {} {\bibfield  {journal} {\bibinfo  {journal} {Phys. Rev. Lett.}\
  }\textbf {\bibinfo {volume} {90}},\ \bibinfo {pages} {247201} (\bibinfo
  {year} {2003})}\BibitemShut {NoStop}%
\bibitem [{\citenamefont {Koopmans}\ \emph {et~al.}(2009)\citenamefont
  {Koopmans}, \citenamefont {Malinowski}, \citenamefont {Dalla~Longa},
  \citenamefont {Steiauf}, \citenamefont {F{\"a}hnle}, \citenamefont {Roth},
  \citenamefont {Cinchetti},\ and\ \citenamefont
  {Aeschlimann}}]{Koopmans:2009ul}%
  \BibitemOpen
  \bibfield  {author} {\bibinfo {author} {\bibfnamefont {B.}~\bibnamefont
  {Koopmans}}, \bibinfo {author} {\bibfnamefont {G.}~\bibnamefont
  {Malinowski}}, \bibinfo {author} {\bibfnamefont {F.}~\bibnamefont
  {Dalla~Longa}}, \bibinfo {author} {\bibfnamefont {D.}~\bibnamefont
  {Steiauf}}, \bibinfo {author} {\bibfnamefont {M.}~\bibnamefont {F{\"a}hnle}},
  \bibinfo {author} {\bibfnamefont {T.}~\bibnamefont {Roth}}, \bibinfo {author}
  {\bibfnamefont {M.}~\bibnamefont {Cinchetti}}, \ and\ \bibinfo {author}
  {\bibfnamefont {M.}~\bibnamefont {Aeschlimann}},\ }\href@noop {} {\bibfield
  {journal} {\bibinfo  {journal} {Nature Mater.}\ }\textbf {\bibinfo {volume}
  {9}},\ \bibinfo {pages} {259} (\bibinfo {year} {2009})}\BibitemShut {NoStop}%
\bibitem [{\citenamefont {Stamm}\ \emph {et~al.}(2007)\citenamefont {Stamm},
  \citenamefont {Kachel}, \citenamefont {Pontius}, \citenamefont {Mitzner},
  \citenamefont {Quast}, \citenamefont {Holldack}, \citenamefont {Khan},
  \citenamefont {Lupulescu}, \citenamefont {Aziz}, \citenamefont {Wietstruk},
  \citenamefont {D{\"u}rr},\ and\ \citenamefont {Eberhardt}}]{Stamm:2007hy}%
  \BibitemOpen
  \bibfield  {author} {\bibinfo {author} {\bibfnamefont {C.}~\bibnamefont
  {Stamm}}, \bibinfo {author} {\bibfnamefont {T.}~\bibnamefont {Kachel}},
  \bibinfo {author} {\bibfnamefont {N.}~\bibnamefont {Pontius}}, \bibinfo
  {author} {\bibfnamefont {R.}~\bibnamefont {Mitzner}}, \bibinfo {author}
  {\bibfnamefont {T.}~\bibnamefont {Quast}}, \bibinfo {author} {\bibfnamefont
  {K.}~\bibnamefont {Holldack}}, \bibinfo {author} {\bibfnamefont
  {S.}~\bibnamefont {Khan}}, \bibinfo {author} {\bibfnamefont {C.}~\bibnamefont
  {Lupulescu}}, \bibinfo {author} {\bibfnamefont {E.~F.}\ \bibnamefont {Aziz}},
  \bibinfo {author} {\bibfnamefont {M.}~\bibnamefont {Wietstruk}}, \bibinfo
  {author} {\bibfnamefont {H.~A.}\ \bibnamefont {D{\"u}rr}}, \ and\ \bibinfo
  {author} {\bibfnamefont {W.}~\bibnamefont {Eberhardt}},\ }\href@noop {}
  {\bibfield  {journal} {\bibinfo  {journal} {Nature Mater.}\ }\textbf
  {\bibinfo {volume} {6}},\ \bibinfo {pages} {740} (\bibinfo {year}
  {2007})}\BibitemShut {NoStop}%
\bibitem [{\citenamefont {Ju}\ \emph {et~al.}(2004)\citenamefont {Ju},
  \citenamefont {Hohlfeld}, \citenamefont {Bergman}, \citenamefont {van~de
  Veerdonk}, \citenamefont {Mryasov}, \citenamefont {Kim}, \citenamefont {Wu},
  \citenamefont {Weller},\ and\ \citenamefont {Koopmans}}]{Ju:2004jw}%
  \BibitemOpen
  \bibfield  {author} {\bibinfo {author} {\bibfnamefont {G.}~\bibnamefont
  {Ju}}, \bibinfo {author} {\bibfnamefont {J.}~\bibnamefont {Hohlfeld}},
  \bibinfo {author} {\bibfnamefont {B.}~\bibnamefont {Bergman}}, \bibinfo
  {author} {\bibfnamefont {R.~J.~M.}~\bibnamefont {van~de Veerdonk}}, \bibinfo
  {author} {\bibfnamefont {O.~N.}~\bibnamefont {Mryasov}}, \bibinfo {author}
  {\bibfnamefont {J.-Y.}\ \bibnamefont {Kim}}, \bibinfo {author} {\bibfnamefont
  {X.}~\bibnamefont {Wu}}, \bibinfo {author} {\bibfnamefont {D.}~\bibnamefont
  {Weller}}, \ and\ \bibinfo {author} {\bibfnamefont {B.}~\bibnamefont
  {Koopmans}},\ }\href@noop {} {\bibfield  {journal} {\bibinfo  {journal}
  {Phys. Rev. Lett.}\ }\textbf {\bibinfo {volume} {93}},\ \bibinfo {pages} {197403} (\bibinfo {year}
  {2004})}\BibitemShut {NoStop}%
\bibitem [{\citenamefont {Radu}\ \emph {et~al.}(2010)\citenamefont {Radu},
  \citenamefont {Stamm}, \citenamefont {Pontius}, \citenamefont {Kachel},
  \citenamefont {Ramm}, \citenamefont {Thiele}, \citenamefont {D{\"u}rr},\ and\
  \citenamefont {Back}}]{Radu:2010ex}%
  \BibitemOpen
  \bibfield  {author} {\bibinfo {author} {\bibfnamefont {I.}~\bibnamefont
  {Radu}}, \bibinfo {author} {\bibfnamefont {C.}~\bibnamefont {Stamm}},
  \bibinfo {author} {\bibfnamefont {N.}~\bibnamefont {Pontius}}, \bibinfo
  {author} {\bibfnamefont {T.}~\bibnamefont {Kachel}}, \bibinfo {author}
  {\bibfnamefont {P.}~\bibnamefont {Ramm}}, \bibinfo {author} {\bibfnamefont
  {J.-U.}\ \bibnamefont {Thiele}}, \bibinfo {author} {\bibfnamefont {H.~A.}\
  \bibnamefont {D{\"u}rr}}, \ and\ \bibinfo {author} {\bibfnamefont {C.~H.}\
  \bibnamefont {Back}},\ }\href@noop {} {\bibfield  {journal} {\bibinfo
  {journal} {Phys. Rev. B}\ }\textbf {\bibinfo {volume} {81}},\ \bibinfo {pages} {104415} (\bibinfo {year}
  {2010})}\BibitemShut {NoStop}%
\bibitem [{\citenamefont {Radu}\ \emph {et~al.}(2011)\citenamefont {Radu},
  \citenamefont {Vahaplar}, \citenamefont {Stamm}, \citenamefont {Kachel},
  \citenamefont {Pontius}, \citenamefont {D{\"u}rr}, \citenamefont {Ostler},
  \citenamefont {Barker}, \citenamefont {Evans}, \citenamefont {Chantrell},
  \citenamefont {Tsukamoto}, \citenamefont {Itoh}, \citenamefont {Kirilyuk},
  \citenamefont {Rasing},\ and\ \citenamefont {Kimel}}]{Radu:2011kr}%
  \BibitemOpen
  \bibfield  {author} {\bibinfo {author} {\bibfnamefont {I.}~\bibnamefont
  {Radu}}, \bibinfo {author} {\bibfnamefont {K.}~\bibnamefont {Vahaplar}},
  \bibinfo {author} {\bibfnamefont {C.}~\bibnamefont {Stamm}}, \bibinfo
  {author} {\bibfnamefont {T.}~\bibnamefont {Kachel}}, \bibinfo {author}
  {\bibfnamefont {N.}~\bibnamefont {Pontius}}, \bibinfo {author} {\bibfnamefont
  {H.~A.}\ \bibnamefont {D{\"u}rr}}, \bibinfo {author} {\bibfnamefont {T.~A.}\
  \bibnamefont {Ostler}}, \bibinfo {author} {\bibfnamefont {J.}~\bibnamefont
  {Barker}}, \bibinfo {author} {\bibfnamefont {R.~F.~L.}\ \bibnamefont
  {Evans}}, \bibinfo {author} {\bibfnamefont {R.~W.}\ \bibnamefont
  {Chantrell}}, \bibinfo {author} {\bibfnamefont {A.}~\bibnamefont
  {Tsukamoto}}, \bibinfo {author} {\bibfnamefont {A.}~\bibnamefont {Itoh}},
  \bibinfo {author} {\bibfnamefont {A.}~\bibnamefont {Kirilyuk}}, \bibinfo
  {author} {\bibfnamefont {T.}~\bibnamefont {Rasing}}, \ and\ \bibinfo {author}
  {\bibfnamefont {A.~V.}\ \bibnamefont {Kimel}},\ }\href@noop {} {\bibfield
  {journal} {\bibinfo  {journal} {Nature}\ }\textbf {\bibinfo {volume} {472}},\
  \bibinfo {pages} {205} (\bibinfo {year} {2011})}\BibitemShut {NoStop}%
\bibitem [{\citenamefont {Seu}\ \emph {et~al.}(2010)\citenamefont {Seu},
  \citenamefont {Roy}, \citenamefont {Turner}, \citenamefont {Park},
  \citenamefont {Falco},\ and\ \citenamefont {Kevan}}]{Seu:2010ec}%
  \BibitemOpen
  \bibfield  {author} {\bibinfo {author} {\bibfnamefont {K.~A.}\ \bibnamefont
  {Seu}}, \bibinfo {author} {\bibfnamefont {S.}~\bibnamefont {Roy}}, \bibinfo
  {author} {\bibfnamefont {J.~J.}\ \bibnamefont {Turner}}, \bibinfo {author}
  {\bibfnamefont {S.}~\bibnamefont {Park}}, \bibinfo {author} {\bibfnamefont
  {C.~M.}\ \bibnamefont {Falco}}, \ and\ \bibinfo {author} {\bibfnamefont
  {S.~D.}\ \bibnamefont {Kevan}},\ }\href@noop {} {\bibfield  {journal}
  {\bibinfo  {journal} {Phys. Rev. B}\ }\textbf {\bibinfo {volume} {82}},\
  \bibinfo {pages} {012404} (\bibinfo {year} {2010})}\BibitemShut {NoStop}%
\bibitem [{\citenamefont {Cheong}\ and\ \citenamefont
  {Mostovoy}(2007)}]{Cheong:2007gd}%
  \BibitemOpen
  \bibfield  {author} {\bibinfo {author} {\bibfnamefont {S.-W.}\ \bibnamefont
  {Cheong}}\ and\ \bibinfo {author} {\bibfnamefont {M.}~\bibnamefont
  {Mostovoy}},\ }\href@noop {} {\bibfield  {journal} {\bibinfo  {journal}
  {Nature Mater.}\ }\textbf {\bibinfo {volume} {6}},\ \bibinfo {pages} {13}
  (\bibinfo {year} {2007})}\BibitemShut {NoStop}%
\bibitem [{\citenamefont {{\AA}sbrink}\ and\ \citenamefont
  {Norrby}(1970)}]{Asbrink:1970vb}%
  \BibitemOpen
  \bibfield  {author} {\bibinfo {author} {\bibfnamefont {S.}~\bibnamefont
  {{\AA}sbrink}}\ and\ \bibinfo {author} {\bibfnamefont {L.-J.}\ \bibnamefont
  {Norrby}},\ }\href@noop {} {\bibfield  {journal} {\bibinfo  {journal} {Acta
  Cryst. B}\ }\textbf {\bibinfo {volume} {26}},\ \bibinfo {pages} {8} (\bibinfo
  {year} {1970})}\BibitemShut {NoStop}%
\bibitem [{\citenamefont {Forsyth}\ \emph {et~al.}(1988)\citenamefont
  {Forsyth}, \citenamefont {Brown},\ and\ \citenamefont
  {Wanklyn}}]{Forsyth:1988tv}%
  \BibitemOpen
  \bibfield  {author} {\bibinfo {author} {\bibfnamefont {J.~B.}\ \bibnamefont
  {Forsyth}}, \bibinfo {author} {\bibfnamefont {P.~J.}\ \bibnamefont {Brown}},
  \ and\ \bibinfo {author} {\bibfnamefont {B.~M.}\ \bibnamefont {Wanklyn}},\
  }\href@noop {} {\bibfield  {journal} {\bibinfo  {journal} {J. Phys. C}\
  }\textbf {\bibinfo {volume} {21}},\ \bibinfo {pages} {2917} (\bibinfo {year}
  {1988})}\BibitemShut {NoStop}%
\bibitem [{\citenamefont {Yang}\ \emph {et~al.}(1989)\citenamefont {Yang},
  \citenamefont {Thurston}, \citenamefont {Tranquada},\ and\ \citenamefont
  {Shirane}}]{Yang:1989uy}%
  \BibitemOpen
  \bibfield  {author} {\bibinfo {author} {\bibfnamefont {B.~X.}\ \bibnamefont
  {Yang}}, \bibinfo {author} {\bibfnamefont {T.~R.}\ \bibnamefont {Thurston}},
  \bibinfo {author} {\bibfnamefont {J.~M.}\ \bibnamefont {Tranquada}}, \ and\
  \bibinfo {author} {\bibfnamefont {G.}~\bibnamefont {Shirane}},\ }\href@noop
  {} {\bibfield  {journal} {\bibinfo  {journal} {Phys. Rev. B}\ }\textbf
  {\bibinfo {volume} {39}},\ \bibinfo {pages} {4343} (\bibinfo {year}
  {1989})}\BibitemShut {NoStop}%
\bibitem [{\citenamefont {Brown}\ \emph {et~al.}(1991)\citenamefont {Brown},
  \citenamefont {Chattopadhyay}, \citenamefont {Forsyth}, \citenamefont
  {Nunez},\ and\ \citenamefont {Tasset}}]{Brown:1991uz}%
  \BibitemOpen
  \bibfield  {author} {\bibinfo {author} {\bibfnamefont {P.~J.}\ \bibnamefont
  {Brown}}, \bibinfo {author} {\bibfnamefont {T.}~\bibnamefont
  {Chattopadhyay}}, \bibinfo {author} {\bibfnamefont {J.~B.}\ \bibnamefont
  {Forsyth}}, \bibinfo {author} {\bibfnamefont {V.}~\bibnamefont {Nunez}}, \
  and\ \bibinfo {author} {\bibfnamefont {F.}~\bibnamefont {Tasset}},\
  }\href@noop {} {\bibfield  {journal} {\bibinfo  {journal} {J. Phys. Cond.
  Mat. J Phys Cond Mat}\ }\textbf {\bibinfo {volume} {3}},\ \bibinfo {pages}
  {4281} (\bibinfo {year} {1991})}\BibitemShut {NoStop}%
\bibitem [{\citenamefont {Kimura}\ \emph {et~al.}(2008)\citenamefont {Kimura},
  \citenamefont {Sekio}, \citenamefont {Nakamura}, \citenamefont {Siegrist},\
  and\ \citenamefont {Ramirez}}]{Kimura:2008ks}%
  \BibitemOpen
  \bibfield  {author} {\bibinfo {author} {\bibfnamefont {T.}~\bibnamefont
  {Kimura}}, \bibinfo {author} {\bibfnamefont {Y.}~\bibnamefont {Sekio}},
  \bibinfo {author} {\bibfnamefont {H.}~\bibnamefont {Nakamura}}, \bibinfo
  {author} {\bibfnamefont {T.}~\bibnamefont {Siegrist}}, \ and\ \bibinfo
  {author} {\bibfnamefont {A.~P.}\ \bibnamefont {Ramirez}},\ }\href@noop {}
  {\bibfield  {journal} {\bibinfo  {journal} {Nature Mater.}\ }\textbf
  {\bibinfo {volume} {7}},\ \bibinfo {pages} {291} (\bibinfo {year}
  {2008})}\BibitemShut {NoStop}%
\bibitem [{\citenamefont {A{\"\i}n}\ \emph {et~al.}(1989)\citenamefont
  {A{\"\i}n}, \citenamefont {Reichardt}, \citenamefont {Hennion}, \citenamefont
  {Pepy},\ and\ \citenamefont {Wanklyn}}]{Ain:1989uk}%
  \BibitemOpen
  \bibfield  {author} {\bibinfo {author} {\bibfnamefont {M.}~\bibnamefont
  {A{\"\i}n}}, \bibinfo {author} {\bibfnamefont {W.}~\bibnamefont {Reichardt}},
  \bibinfo {author} {\bibfnamefont {B.}~\bibnamefont {Hennion}}, \bibinfo
  {author} {\bibfnamefont {G.}~\bibnamefont {Pepy}}, \ and\ \bibinfo {author}
  {\bibfnamefont {B.~M.}\ \bibnamefont {Wanklyn}},\ }\href@noop {} {\bibfield
  {journal} {\bibinfo  {journal} {Physica C}\ }\textbf {\bibinfo {volume}
  {162}},\ \bibinfo {pages} {1279} (\bibinfo {year} {1989})}\BibitemShut
  {NoStop}%
\bibitem [{\citenamefont {Boothroyd}\ \emph {et~al.}(1997)\citenamefont
  {Boothroyd}, \citenamefont {Mukherjee}, \citenamefont {Fulton}, \citenamefont
  {Perring}, \citenamefont {Eccleston}, \citenamefont {Mook},\ and\
  \citenamefont {Wanklyn}}]{Boothroyd:1997wi}%
  \BibitemOpen
  \bibfield  {author} {\bibinfo {author} {\bibfnamefont {A.~T.}\ \bibnamefont
  {Boothroyd}}, \bibinfo {author} {\bibfnamefont {A.}~\bibnamefont
  {Mukherjee}}, \bibinfo {author} {\bibfnamefont {S.}~\bibnamefont {Fulton}},
  \bibinfo {author} {\bibfnamefont {T.~G.}\ \bibnamefont {Perring}}, \bibinfo
  {author} {\bibfnamefont {R.~S.}\ \bibnamefont {Eccleston}}, \bibinfo {author}
  {\bibfnamefont {H.~A.}\ \bibnamefont {Mook}}, \ and\ \bibinfo {author}
  {\bibfnamefont {B.~M.}\ \bibnamefont {Wanklyn}},\ }\href@noop {} {\bibfield
  {journal} {\bibinfo  {journal} {Physica B}\ }\textbf {\bibinfo {volume}
  {234}},\ \bibinfo {pages} {731} (\bibinfo {year} {1997})}\BibitemShut
  {NoStop}%
\bibitem [{\citenamefont {Yablonskii}(1990)}]{Yablonskii:1990te}%
  \BibitemOpen
  \bibfield  {author} {\bibinfo {author} {\bibfnamefont {D.~A.}\ \bibnamefont
  {Yablonskii}},\ }\href@noop {} {\bibfield  {journal} {\bibinfo  {journal}
  {Physica C}\ }\textbf {\bibinfo {volume} {171}},\ \bibinfo {pages} {454}
  (\bibinfo {year} {1990})}\BibitemShut {NoStop}%
\bibitem [{\citenamefont {Sukhorukov}\ \emph {et~al.}(1995)\citenamefont
  {Sukhorukov}, \citenamefont {Loshkareva}, \citenamefont {Moskvin},\ and\
  \citenamefont {Samokhvalov}}]{Sukhorukov:1995ts}%
  \BibitemOpen
  \bibfield  {author} {\bibinfo {author} {\bibfnamefont {Y.~P.}\ \bibnamefont
  {Sukhorukov}}, \bibinfo {author} {\bibfnamefont {N.~N.}\ \bibnamefont
  {Loshkareva}}, \bibinfo {author} {\bibfnamefont {A.~S.}\ \bibnamefont
  {Moskvin}}, \ and\ \bibinfo {author} {\bibfnamefont {A.~A.}\ \bibnamefont
  {Samokhvalov}},\ }\href@noop {} {\bibfield  {journal} {\bibinfo  {journal}
  {Zh. Eksp. Teor. Fiz.}\ }\textbf {\bibinfo {volume} {108}},\ \bibinfo {pages}
  {1821} (\bibinfo {year} {1995})}\BibitemShut {NoStop}%
\bibitem [{\citenamefont {Scagnoli}\ \emph {et~al.}(2011)\citenamefont
  {Scagnoli}, \citenamefont {Staub}, \citenamefont {Bodenthin}, \citenamefont
  {de~Souza}, \citenamefont {Garc{\'\i}a-Fern{\'a}ndez}, \citenamefont
  {Garganourakis}, \citenamefont {Boothroyd}, \citenamefont {Prabhakaran},\
  and\ \citenamefont {Lovesey}}]{Scagnoli:2011db}%
  \BibitemOpen
  \bibfield  {author} {\bibinfo {author} {\bibfnamefont {V.}~\bibnamefont
  {Scagnoli}}, \bibinfo {author} {\bibfnamefont {U.}~\bibnamefont {Staub}},
  \bibinfo {author} {\bibfnamefont {Y.}~\bibnamefont {Bodenthin}}, \bibinfo
  {author} {\bibfnamefont {R.~A.}\ \bibnamefont {de~Souza}}, \bibinfo {author}
  {\bibfnamefont {M.}~\bibnamefont {Garc{\'\i}a-Fern{\'a}ndez}}, \bibinfo
  {author} {\bibfnamefont {M.}~\bibnamefont {Garganourakis}}, \bibinfo {author}
  {\bibfnamefont {A.~T.}\ \bibnamefont {Boothroyd}}, \bibinfo {author}
  {\bibfnamefont {D.}~\bibnamefont {Prabhakaran}}, \ and\ \bibinfo {author}
  {\bibfnamefont {S.~W.}\ \bibnamefont {Lovesey}},\ }\href@noop {} {\bibfield
  {journal} {\bibinfo  {journal} {Science}\ }\textbf {\bibinfo {volume}
  {332}},\ \bibinfo {pages} {696} (\bibinfo {year} {2011})}\BibitemShut
  {NoStop}%
\bibitem [{\citenamefont {Wu}\ \emph {et~al.}(2010)\citenamefont {Wu},
  \citenamefont {Huang}, \citenamefont {Okamoto}, \citenamefont {Huang},
  \citenamefont {Sekio}, \citenamefont {Kimura},\ and\ \citenamefont
  {Chen}}]{Wu:2010cf}%
  \BibitemOpen
  \bibfield  {author} {\bibinfo {author} {\bibfnamefont {W.~B.}\ \bibnamefont
  {Wu}}, \bibinfo {author} {\bibfnamefont {D.~J.}\ \bibnamefont {Huang}},
  \bibinfo {author} {\bibfnamefont {J.}~\bibnamefont {Okamoto}}, \bibinfo
  {author} {\bibfnamefont {S.~W.}\ \bibnamefont {Huang}}, \bibinfo {author}
  {\bibfnamefont {Y.}~\bibnamefont {Sekio}}, \bibinfo {author} {\bibfnamefont
  {T.}~\bibnamefont {Kimura}}, \ and\ \bibinfo {author} {\bibfnamefont {C.~T.}\
  \bibnamefont {Chen}},\ }\href@noop {} {\bibfield  {journal} {\bibinfo
  {journal} {Phys. Rev. B}\ }\textbf {\bibinfo {volume} {81}},\ \bibinfo {pages} 
  {172409} (\bibinfo {year}
  {2010})}\BibitemShut {NoStop}%
\bibitem [{\citenamefont {Varma}(1997)}]{Varma:1997ue}%
  \BibitemOpen
  \bibfield  {author} {\bibinfo {author} {\bibfnamefont {C.~M.}\ \bibnamefont
  {Varma}},\ }\href@noop {} {\bibfield  {journal} {\bibinfo  {journal} {Phys.
  Rev. B}\ }\textbf {\bibinfo {volume} {55}},\ \bibinfo {pages} {14554}
  (\bibinfo {year} {1997})}\BibitemShut {NoStop}%
\bibitem [{\citenamefont {Prabhakaran}\ and\ \citenamefont
  {Boothroyd}(2003)}]{Prabhakaran:2003hd}%
  \BibitemOpen
  \bibfield  {author} {\bibinfo {author} {\bibfnamefont {D.}~\bibnamefont
  {Prabhakaran}}\ and\ \bibinfo {author} {\bibfnamefont {A.~T.}\ \bibnamefont
  {Boothroyd}},\ }\href@noop {} {\bibfield  {journal} {\bibinfo  {journal} {J.
  Cryst. Growth}\ }\textbf {\bibinfo {volume} {250}},\ \bibinfo {pages} {77}
  (\bibinfo {year} {2003})}\BibitemShut {NoStop}%
\bibitem [{\citenamefont {Emma}\ \emph {et~al.}(2010)\citenamefont {Emma},
  \citenamefont {Akre}, \citenamefont {Arthur}, \citenamefont {Bionta},
  \citenamefont {Bostedt}, \citenamefont {Bozek}, \citenamefont {Brachmann},
  \citenamefont {Bucksbaum}, \citenamefont {Coffee}, \citenamefont {Decker},
  \citenamefont {Ding}, \citenamefont {Dowell}, \citenamefont {Edstrom},
  \citenamefont {Fisher}, \citenamefont {Frisch}, \citenamefont {Gilevich},
  \citenamefont {Hastings}, \citenamefont {Hays}, \citenamefont {Hering},
  \citenamefont {Huang}, \citenamefont {Iverson}, \citenamefont {Loos},
  \citenamefont {Messerschmidt}, \citenamefont {Miahnahri}, \citenamefont
  {Moeller}, \citenamefont {Nuhn}, \citenamefont {Pile}, \citenamefont
  {Ratner}, \citenamefont {Rzepiela}, \citenamefont {Schultz}, \citenamefont
  {Smith}, \citenamefont {Stefan}, \citenamefont {Tompkins}, \citenamefont
  {Turner}, \citenamefont {Welch}, \citenamefont {White}, \citenamefont {Wu},
  \citenamefont {Yocky},\ and\ \citenamefont {Galayda}}]{Emma:2010gu}%
  \BibitemOpen
  \bibfield  {author} {\bibinfo {author} {\bibfnamefont {P.}~\bibnamefont
  {Emma}}, \bibinfo {author} {\bibfnamefont {R.}~\bibnamefont {Akre}}, \bibinfo
  {author} {\bibfnamefont {J.}~\bibnamefont {Arthur}}, \bibinfo {author}
  {\bibfnamefont {R.}~\bibnamefont {Bionta}}, \bibinfo {author} {\bibfnamefont
  {C.}~\bibnamefont {Bostedt}}, \bibinfo {author} {\bibfnamefont
  {J.}~\bibnamefont {Bozek}}, \bibinfo {author} {\bibfnamefont
  {A.}~\bibnamefont {Brachmann}}, \bibinfo {author} {\bibfnamefont
  {P.}~\bibnamefont {Bucksbaum}}, \bibinfo {author} {\bibfnamefont
  {R.}~\bibnamefont {Coffee}}, \bibinfo {author} {\bibfnamefont {F.~J.}\
  \bibnamefont {Decker}}, \bibinfo {author} {\bibfnamefont {Y.}~\bibnamefont
  {Ding}}, \bibinfo {author} {\bibfnamefont {D.}~\bibnamefont {Dowell}},
  \bibinfo {author} {\bibfnamefont {S.}~\bibnamefont {Edstrom}}, \bibinfo
  {author} {\bibfnamefont {A.}~\bibnamefont {Fisher}}, \bibinfo {author}
  {\bibfnamefont {J.}~\bibnamefont {Frisch}}, \bibinfo {author} {\bibfnamefont
  {S.}~\bibnamefont {Gilevich}}, \bibinfo {author} {\bibfnamefont
  {J.}~\bibnamefont {Hastings}}, \bibinfo {author} {\bibfnamefont
  {G.}~\bibnamefont {Hays}}, \bibinfo {author} {\bibfnamefont {P.}~\bibnamefont
  {Hering}}, \bibinfo {author} {\bibfnamefont {Z.}~\bibnamefont {Huang}},
  \bibinfo {author} {\bibfnamefont {R.}~\bibnamefont {Iverson}}, \bibinfo
  {author} {\bibfnamefont {H.}~\bibnamefont {Loos}}, \bibinfo {author}
  {\bibfnamefont {M.}~\bibnamefont {Messerschmidt}}, \bibinfo {author}
  {\bibfnamefont {A.}~\bibnamefont {Miahnahri}}, \bibinfo {author}
  {\bibfnamefont {S.}~\bibnamefont {Moeller}}, \bibinfo {author} {\bibfnamefont
  {H.~D.}\ \bibnamefont {Nuhn}}, \bibinfo {author} {\bibfnamefont
  {G.}~\bibnamefont {Pile}}, \bibinfo {author} {\bibfnamefont {D.}~\bibnamefont
  {Ratner}}, \bibinfo {author} {\bibfnamefont {J.}~\bibnamefont {Rzepiela}},
  \bibinfo {author} {\bibfnamefont {D.}~\bibnamefont {Schultz}}, \bibinfo
  {author} {\bibfnamefont {T.}~\bibnamefont {Smith}}, \bibinfo {author}
  {\bibfnamefont {P.}~\bibnamefont {Stefan}}, \bibinfo {author} {\bibfnamefont
  {H.}~\bibnamefont {Tompkins}}, \bibinfo {author} {\bibfnamefont
  {J.}~\bibnamefont {Turner}}, \bibinfo {author} {\bibfnamefont
  {J.}~\bibnamefont {Welch}}, \bibinfo {author} {\bibfnamefont
  {W.}~\bibnamefont {White}}, \bibinfo {author} {\bibfnamefont
  {J.}~\bibnamefont {Wu}}, \bibinfo {author} {\bibfnamefont {G.}~\bibnamefont
  {Yocky}}, \ and\ \bibinfo {author} {\bibfnamefont {J.}~\bibnamefont
  {Galayda}},\ }\href@noop {} {\bibfield  {journal} {\bibinfo  {journal}
  {Nature Photon.}\ }\textbf {\bibinfo {volume} {4}},\ \bibinfo {pages} {641}
  (\bibinfo {year} {2010})}\BibitemShut {NoStop}%
\bibitem [{\citenamefont {Johnson}\ \emph {et~al.}(2009)\citenamefont
  {Johnson}, \citenamefont {Beaud}, \citenamefont {Vorobeva}, \citenamefont
  {Milne}, \citenamefont {Murray}, \citenamefont {Fahy},\ and\ \citenamefont
  {Ingold}}]{Johnson:2009ty}%
  \BibitemOpen
  \bibfield  {author} {\bibinfo {author} {\bibfnamefont {S.~L.}\ \bibnamefont
  {Johnson}}, \bibinfo {author} {\bibfnamefont {P.}~\bibnamefont {Beaud}},
  \bibinfo {author} {\bibfnamefont {E.}~\bibnamefont {Vorobeva}}, \bibinfo
  {author} {\bibfnamefont {C.~J.}\ \bibnamefont {Milne}}, \bibinfo {author}
  {\bibfnamefont {{\'E}.~D.}\ \bibnamefont {Murray}}, \bibinfo {author}
  {\bibfnamefont {S.}~\bibnamefont {Fahy}}, \ and\ \bibinfo {author}
  {\bibfnamefont {G.}~\bibnamefont {Ingold}},\ }\href@noop {} {\bibfield
  {journal} {\bibinfo  {journal} {Phys. Rev. Lett.}\ }\textbf {\bibinfo
  {volume} {102}},\ \bibinfo {pages} {175503} (\bibinfo {year}
  {2009})}\BibitemShut {NoStop}%
\bibitem [{\citenamefont {Johnson}\ \emph {et~al.}(2010)\citenamefont
  {Johnson}, \citenamefont {Beaud}, \citenamefont {Vorobeva}, \citenamefont
  {Milne}, \citenamefont {Murray}, \citenamefont {Fahy},\ and\ \citenamefont
  {Ingold}}]{Johnson:2010em}%
  \BibitemOpen
  \bibfield  {author} {\bibinfo {author} {\bibfnamefont {S.~L.}\ \bibnamefont
  {Johnson}}, \bibinfo {author} {\bibfnamefont {P.}~\bibnamefont {Beaud}},
  \bibinfo {author} {\bibfnamefont {E.}~\bibnamefont {Vorobeva}}, \bibinfo
  {author} {\bibfnamefont {C.~J.}\ \bibnamefont {Milne}}, \bibinfo {author}
  {\bibfnamefont {{\'E}.~D.}\ \bibnamefont {Murray}}, \bibinfo {author}
  {\bibfnamefont {S.}~\bibnamefont {Fahy}}, \ and\ \bibinfo {author}
  {\bibfnamefont {G.}~\bibnamefont {Ingold}},\ }\href@noop {} {\bibfield
  {journal} {\bibinfo  {journal} {Acta Cryst. A}\ }\textbf {\bibinfo {volume}
  {66}},\ \bibinfo {pages} {157} (\bibinfo {year} {2010})}\BibitemShut
  {NoStop}%
\bibitem [{EPA()}]{EPAPS}%
  \BibitemOpen
  \href@noop {} {}\bibinfo {note} {See supplementary information figures, included at the end of this document.}\BibitemShut {Stop}%
\bibitem [{\citenamefont {Cavalleri}\ \emph {et~al.}(2004)\citenamefont
  {Cavalleri}, \citenamefont {Dekorsy}, \citenamefont {Chong}, \citenamefont
  {Kieffer},\ and\ \citenamefont {Schoenlein}}]{Cavalleri:2004eh}%
  \BibitemOpen
  \bibfield  {author} {\bibinfo {author} {\bibfnamefont {A.}~\bibnamefont
  {Cavalleri}}, \bibinfo {author} {\bibfnamefont {Th.}~\bibnamefont {Dekorsy}},
  \bibinfo {author} {\bibfnamefont {H.~H.~W.}~\bibnamefont {Chong}}, \bibinfo
  {author} {\bibfnamefont {J.~C.}~\bibnamefont {Kieffer}}, \ and\ \bibinfo
  {author} {\bibfnamefont {R.~W.}~\bibnamefont {Schoenlein}},\ }\href@noop {}
  {\bibfield  {journal} {\bibinfo  {journal} {Phys. Rev. B}\ }\textbf {\bibinfo
  {volume} {70}},\ \bibinfo {pages} {161102} (\bibinfo {year} {2004})}\BibitemShut {NoStop}%
\bibitem [{\citenamefont {Macrae}\ \emph {et~al.}(2006)\citenamefont {Macrae},
  \citenamefont {Edgington}, \citenamefont {McCabe}, \citenamefont {Pidcock},
  \citenamefont {Shields}, \citenamefont {Taylor}, \citenamefont {Towler},\
  and\ \citenamefont {Streek}}]{Macrae:2006tz}%
  \BibitemOpen
  \bibfield  {author} {\bibinfo {author} {\bibfnamefont {C.~F.}\ \bibnamefont
  {Macrae}}, \bibinfo {author} {\bibfnamefont {P.~R.}\ \bibnamefont
  {Edgington}}, \bibinfo {author} {\bibfnamefont {P.}~\bibnamefont {McCabe}},
  \bibinfo {author} {\bibfnamefont {E.}~\bibnamefont {Pidcock}}, \bibinfo
  {author} {\bibfnamefont {G.~P.}\ \bibnamefont {Shields}}, \bibinfo {author}
  {\bibfnamefont {R.}~\bibnamefont {Taylor}}, \bibinfo {author} {\bibfnamefont
  {M.}~\bibnamefont {Towler}}, \ and\ \bibinfo {author} {\bibfnamefont
  {J.~v.~d.}\ \bibnamefont {Streek}},\ }\href@noop {} {\bibfield  {journal}
  {\bibinfo  {journal} {J. Appl. Cryst.}\ }\textbf {\bibinfo {volume} {39}},\
  \bibinfo {pages} {453} (\bibinfo {year} {2006})}\BibitemShut {NoStop}%
\end{thebibliography}

\clearpage

\begin{figure}
\includegraphics{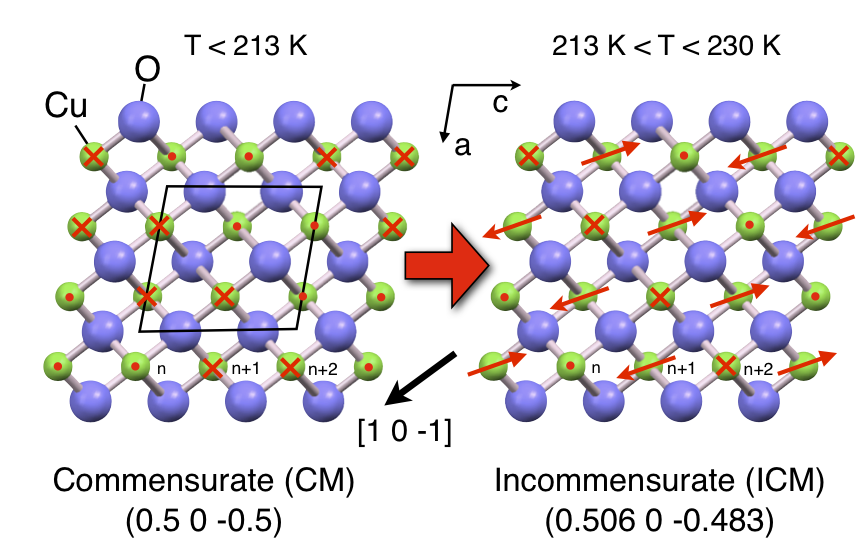}
\caption{
Magnetic structures in CuO, projected onto the (0 1 0) plane~\cite{Asbrink:1970vb,Macrae:2006tz}.  Left: the magnetic structure of the CM phase; right: that of the ICM phase.  Red arrows indicate the orientation of the magnetic dipoles.  }
\end{figure}


\begin{figure}
\includegraphics{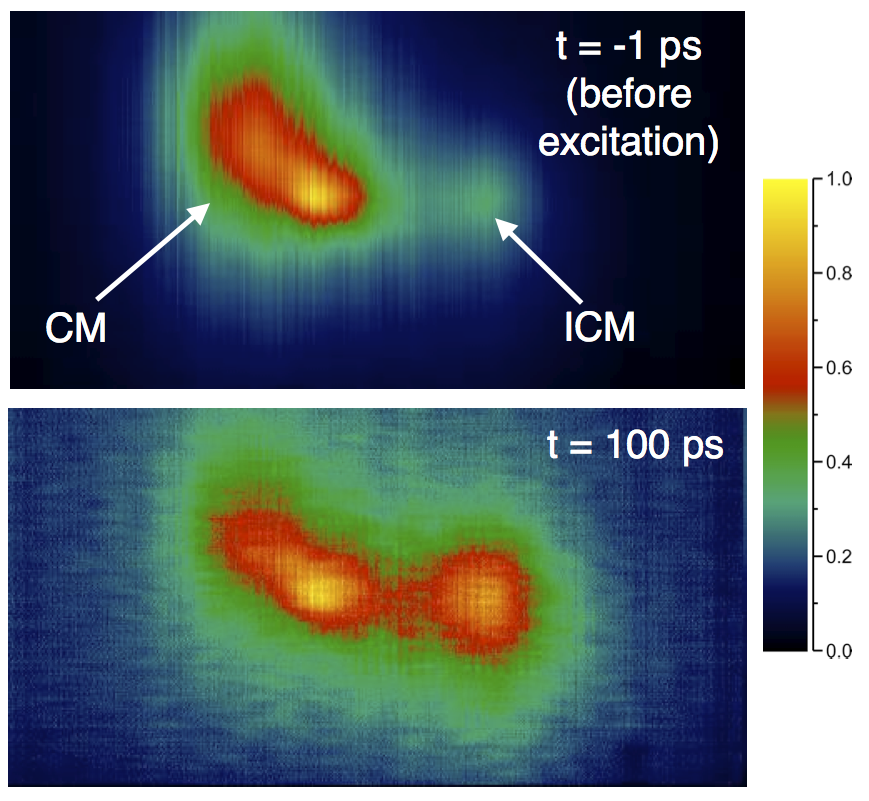}
\caption{
Images of resonant x-ray diffraction.  Upper panel: 1~ps before excitation.  Lower panel: 100~ps after excitation with a femtosecond laser pulse at an incident fluence of 39~mJ/cm$^2$.  The false-color scale for each image is normalized to better show the relative intensities.}
\end{figure}

\begin{figure}
\includegraphics{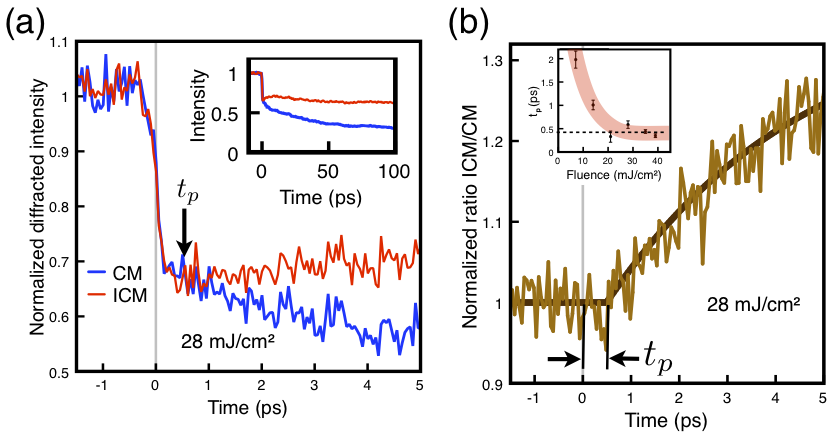}
\caption{(a) Time-dependent normalized diffracted intensity for an excitation fluence of 28~mJ/cm$^2$.  Inset: longer time-scale behavior.  (b) Relative phase population.  The curve is a fit described in the text.  
Inset: dependence of the onset time $t_p$ on the excitation fluence.  The curve is a guide to the eye.  The dashed line shows the time for a 1/4 period of a long-wavelength magnetic excitation in the CM phase (400~fs).  See~\cite{EPAPS} for additional data.}
\end{figure}

\begin{figure}
\includegraphics{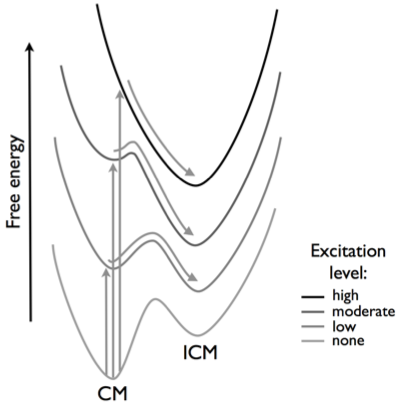}
\caption{
Schematic of the transition.  The ordinate is free energy, and the abscissa is a configurational coordinate.  Stronger excitation causes both a decrease in the relative energy difference between the ICM and CM phases and a decrease in the height of the energy barrier.  At high excitation the change in configurational coordinate is a $1/4$ cycle of a low momentum spin wave.
}
\end{figure}

\clearpage

\includegraphics{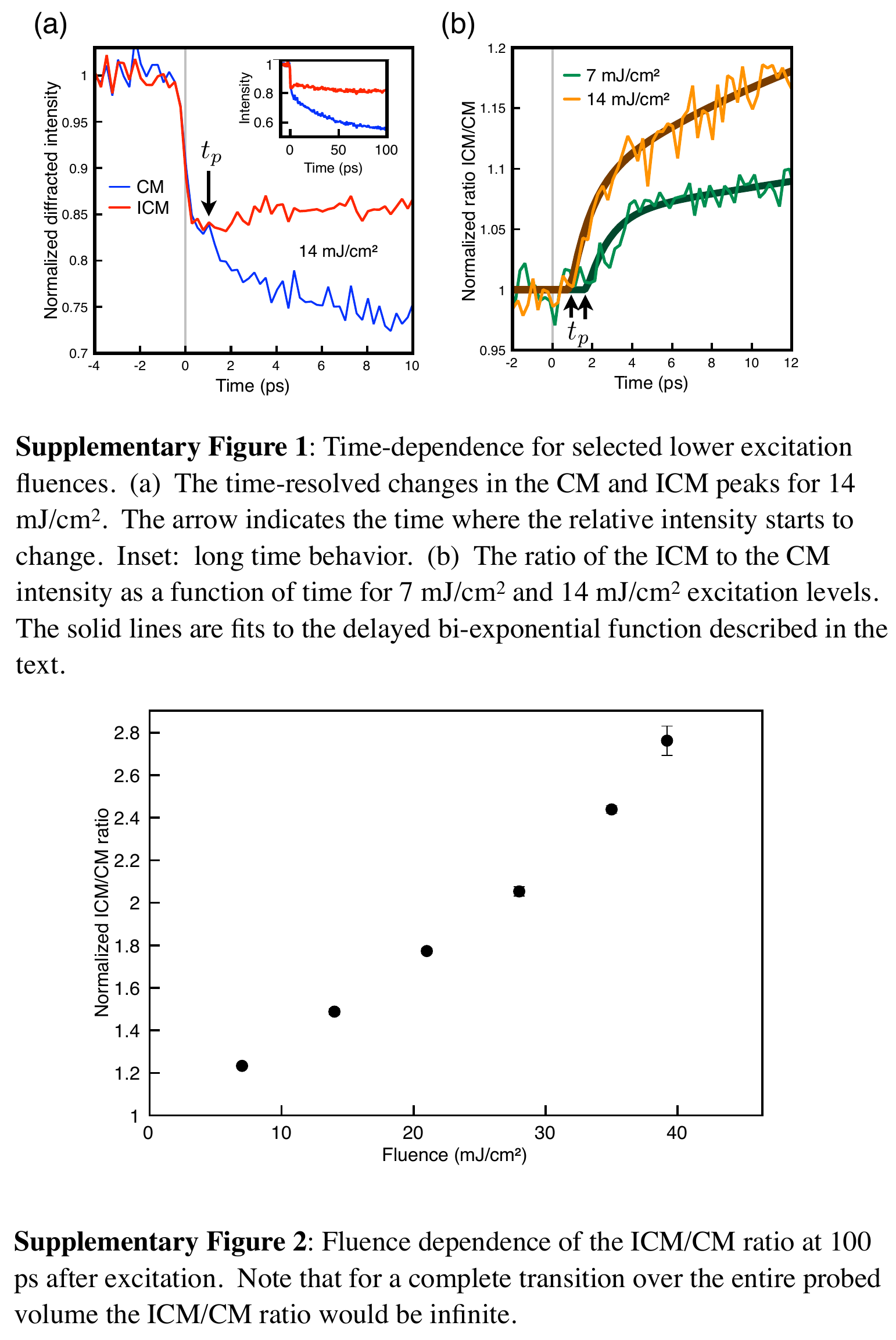}
\includegraphics{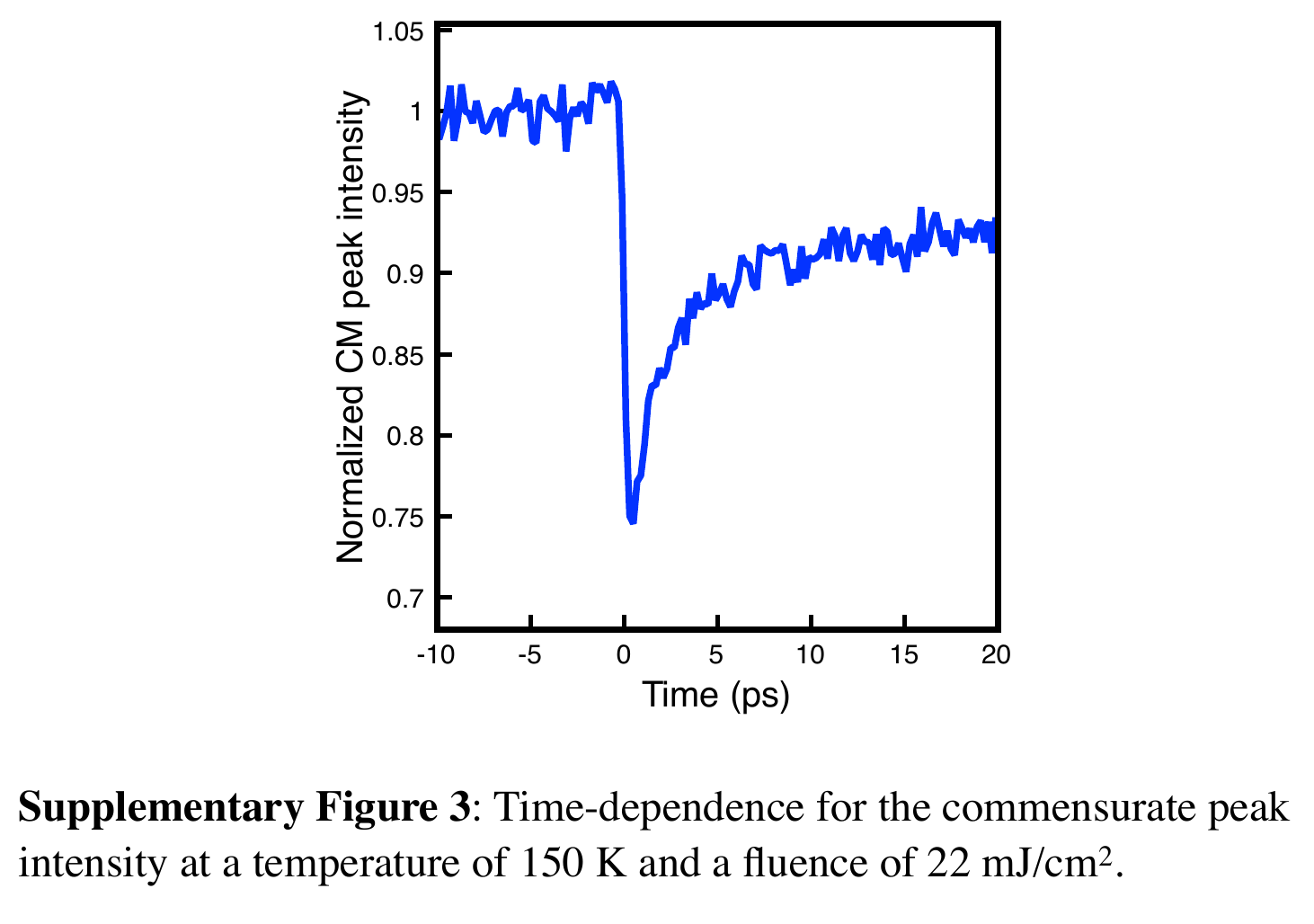}
\end{document}